\documentclass[aps,preprint,nofootinbib]{revtex4}
\usepackage{graphicx}
\usepackage{amsmath,amssymb}
\usepackage{url}
\usepackage{color}
\usepackage{epstopdf}
\newcommand{\lsim}{\mathrel{\mathop{\kern 0pt \rlap
  {\raise.2ex\hbox{$<$}}}
  \lower.9ex\hbox{\kern-.190em $\sim$}}}
\newcommand{\gsim}{\mathrel{\mathop{\kern 0pt \rlap
  {\raise.2ex\hbox{$>$}}}
  \lower.9ex\hbox{\kern-.190em $\sim$}}}

\newcommand{\beq}     {\begin{equation}}
\newcommand{\eeq}     {\end{equation}}
\newcommand{\bea}     {\begin{eqnarray}}
\newcommand{\eea}     {\end{eqnarray}}

\newcommand{\gev}      {{\,\rm GeV}}
\newcommand{\Lg}      {{\cal L}}

\begin{document}
\title{Collider signatures of the Gauge-Higgs Dark Matter}
\author{Kingman Cheung$^{1,2,3}$, Jeonghyeon Song$^{1}$}

\affiliation{
$^1$Division of Quantum Phases \& Devices, School of Physics, 
Konkuk University, Seoul 143-701, Korea \\
$^2$Department of Physics, National Tsing Hua University, 
Hsinchu 300, Taiwan
\\
$^3$Physics Division, National Center for Theoretical Sciences,
Hsinchu 300, Taiwan
}

\date{\today}

\begin{abstract}
A recently proposed 
$SO(5) \times U(1)$ gauge-Higgs unification model in 
the Randall-Sundrum warped space 
contains a very interesting dark matter 
candidate, the Higgs boson. 
As a part of the fifth component of
a five-dimensional gauge field,
the four-dimensional neutral Higgs boson has odd $H$ parity while
all the other standard model particles have even parity
to all orders in perturbation theory.
Based on $H$ parity conserving
$WWHH$ and $ZZHH$ vertices,
we investigate the collider signatures of Higgs dark
matter production associated 
with a standard model $W$ or $Z$ gauge boson at the LHC and the International Linear Collider (ILC). 
The final state consists
of a $W$ or a $Z$ boson with large missing energy.
We found that the level of the signal cross section
is quite hopeless at the LHC, while we may be able to 
identify it at the ILC with polarized electron beams.
\end{abstract}
\maketitle

\section{Introduction}
The presence of cold dark matter (CDM) in our Universe is now well established
by a number of observational experiments, especially the very precise 
measurement of the cosmic microwave background radiation
in the Wilkinson Microwave Anisotropy Probe (WMAP) experiment \cite{wmap}.
Though the gravitational nature of the dark matter (DM) is established, 
we know almost nothing about the particle nature, except that it is,
to a high extent, electrically neutral.

One of the most appealing and natural CDM particle candidates is the
{\it weakly interacting massive particle}\;\cite{hooper}.  
It is a coincidence
that if the dark matter is produced thermally in the early Universe,
the required annihilation cross section 
times the velocity is about $1$ pb.  
This is exactly
the size of the cross sections that one expects from a weak interaction
process and that would give a large to moderate production at the LHC.
In general, the production of dark matter particles
at the LHC gives rise to
large missing energy.  
Thus, the anticipated signature in the 
final state is high-$p_T$ jets or leptons plus large missing energy.

The most studied dark matter candidate is the
neutralino of the supersymmetric models with $R$ parity conservation\,\cite{Jungman:1995df}.
In this work, we study a different scenario,
the $SO(5) \times U(1)$
gauge-Higgs unification model\,\cite{hosotani08,hosotani09} 
based on
the Randall-Sundrum warped space\,\cite{rs}.  
The dark matter is
the Higgs boson\,\cite{ko}, which is a part of the 
fifth component of a gauge boson field in the model.
In such a 5D model, 
the Higgs boson is the fluctuation mode
of the Aharonov-Bohm phase $\hat\theta_H$ in the extra 
dimension\,\cite{ab-phase}.
It was shown that at the value of $\theta_H = \pm \pi/2$
the effective potential of the Higgs boson is minimized.
Furthermore, the invariance of the effective interactions 
under $H \to -H$ prohibits
triple vertices such as $WWH$, $ZZH$, and $\bar{f}f H$, 
which is true to all orders in perturbation theory.
Thus, the Higgs boson is stable and can be a dark matter candidate.

In this model, the interactions of the Higgs boson
with the $W$, $Z$,
and SM fermions are via 4-point vertices.
The Higgs boson can be thermally produced 
in the early Universe via
$WW,ZZ,f\bar f \to H H$.
Like other DM candidates,
the drop of the annihilation rate 
below the universe expansion rate triggered
the freeze-out of the Higgs boson as CDM.
It was shown in Ref.\,\cite{ko} that 
the Higgs boson mass needs to be
at 70 GeV in order to satisfy the constraint from WMAP. 

Focused on 4-point vertices of $HHW^+W^-$ and $HHZZ$,
we study the collider signatures of this dark matter model.
Our main process is
the production of a pair of the Higgs bosons
associated with a $W$ or $Z$ boson.
The final state consists of charged leptons plus large 
missing energy.
However, the detection of this signal is very challenging.
The signal cross section is generically small due to 
the $2\to 3$
process with the weak coupling.
In addition, only one single observable particle in the final state
provides very limited kinematics, 
which could be used to reduce the SM background.
As shown below,
the signal at the LHC is too small to be useful.
On the other hand, the 
International Linear Collider (ILC) 
with high beam polarization\,\cite{ilc}
can substantially improve the sensitivity to the signal.

\section{Effective Interactions and  Relic Density}
We consider a $SO(5) \times U(1)$ gauge-Higgs
unification model 
in the 5D Randall-Sundrum warped space\,\cite{hosotani08}.
The Higgs boson is the fluctuation mode of the
Aharonov-Bohm phase $\hat{\theta}_H$ along the fifth dimension, 
i.e., $\hat{\theta}_H = \theta_H + H(x) /f_H$.
The effective interactions of the Higgs boson
are 
\begin{equation}
\label{eq:effective:V}
{\cal L} = V_{\rm eff}(\hat{\theta}_H)
- m_W^2(\hat{\theta}_H) W^+_\mu W^{-\mu} 
- \frac{1}{2}m_Z^2(\hat{\theta}_H) Z_\mu Z^\mu
,
\end{equation}
where these mass functions are
$m_W(\hat{\theta}_H) =
\frac{1}{2} g f_H \sin \hat{\theta}_H
$ and 
$m_Z(\hat{\theta}_H) =
\frac{1}{2} g_Z f_H \sin \hat{\theta}_H
$.
Here $g$ is the weak gauge coupling
and $g_Z = g/\cos\theta_W$.
The Higgs effective potential $V_{\rm eff}(\hat{\theta}_H)$
is generated at one loop level.
It is finite and cutoff 
independent as well as leading to finite Higgs boson mass:
the gauge hierachy problem does not arise.

Brief comments on the electroweak symmetry breaking
are in order here.
The electroweak symmetry is preserved at $\theta_H =0,\pi$,
as can be seen in the $W$ and $Z$ boson
masses in Eq.(\ref{eq:mw:mz}).
Hosotani {\it et.al.} showed that the contributions
from the gauge bosons and their KK states do not change
the position of 
the global minimum of $\theta_H$\,\cite{hosotani08}.
However, the large but negative 
contribution from the 5D top quark field turns 
$V_{\rm eff}(\hat{\theta}_H)$ upside down,
leading to its global minimum 
at $\theta_H = \pm \pi/2$.
In this model,
the large Yukawa coupling of the top quark triggers the electroweak symmetry
breaking dynamically, and 
the $W$ and $Z$ gauge bosons and the SM fermions
acquire nonzero masses.

At low energy, this model has the SM particles plus
one neutral Higgs boson.
Since $\Lg_{\rm eff}$ in Eq.(\ref{eq:effective:V})
with $\theta =\pi/2$ is
invariant under $H \to -H$,
a new $H$ parity
emerges under which the Higgs boson has odd parity
while all the other SM particles have even parity.
This $H$ parity protects the stability of the Higgs boson,
making it a CDM candidate, 
and prohibits triple vertices of the Higgs boson
with the SM particles.
Note that the $H$ parity is dynamically generated
unlike the $R$ parity in supersymmetric models.

The low energy behavior of the model depends on only 
two parameters, $f_H$ and $m_H$.
First, $f_H$ is determined by the observed $m_W$ and $m_Z$,
i.e., $f_H \approx 246\gev$.
Second, the Higgs boson mass is in principle
determined by the matter content of the model.
Because of the high model dependence,
the Higgs mass has been treated as an unknown parameter.
The observed relic density of CDM by WMAP
fixes the Higgs boson mass about $70\gev$ \cite{ko}.
Such a low Higgs boson mass does not contradict the
LEP bound of the Higgs boson because of the absence 
of the $ZZH$ coupling.
If we further relax the relic density constraint as not overclosing 
the universe, the Higgs boson mass can be  heavier than $70\gev$
\cite{ko}.

We do not consider the 4-point vertices $HH\bar{f}f$, 
where $f$ is the SM fermion. 
This vertex comes from the Yukawa coupling, which is proportional
to the fermion mass.
Although this gives rise to important implications
on the direct detection rate,
the interaction magnitudes involving $HH\bar{f}f$
are very small,
except for the top quark.
The process $pp \to t \bar{t} HH$
is also very challenging to observe
due to the smaller cross section of the $2\to 4$ process
as well as more 
difficult identification of top quarks.
Therefore, we focus on the 4-point vertices
of $HHWW$ and $HHZZ$:
\begin{equation}
{\cal L} =-\frac{1}{8}g_Z^2 H^2 Z_\mu Z^\mu
- \frac{1}{4} g^2 H^2 W^+_\mu W^{-\mu}.
\end{equation}

\section{Collider Signatures at the LHC}

\begin{figure}[t!]
\centering
\includegraphics[scale=1]{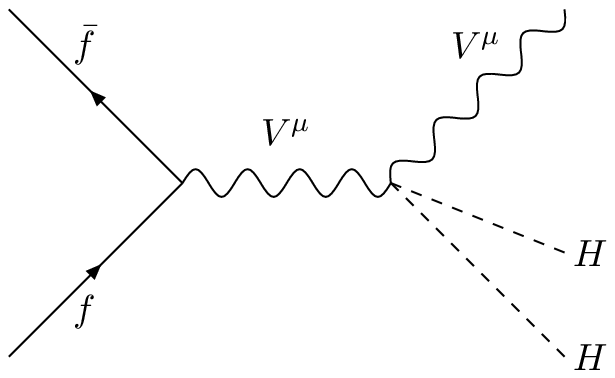}
\caption{\label{fig:feyn} \small 
Feynman diagram for the production of a pair of missing
Higgs boson associated with a SM gauge boson $V$.
}
\end{figure}

As suggested in Ref.\,\cite{Hosotani10},
the collider signature of the Higgs DM involves their
pair production.
Sole pair production via processes such
as $WW \to HH$, $ZZ\to HH$, and $gg \to HH$
leaves only missing energy in the final state, which
has nothing to be triggered on.
We need at least one visible particle to probe the missing 
transverse energy.

We consider the following production of a pair of Higgs bosons
associated with 
a SM gauge boson $V=W,Z$ at the LHC and ILC:
$
  q (p_1) +\bar q(p_2)  \to
 V(q_1)+ H (k_1)+H(k_2)
$.
This process is basically through the
Feynman diagram in Fig.\,\ref{fig:feyn}.
The first subprocess that we consider at the LHC  is 
$q\bar{q} \to ZHH$.
The spin- and color-averaged matrix element squared is given by
\begin{equation}
\overline{\left| M \right |^2}
(q\bar{q} \to ZHH)= \frac{1}{4}\,\frac{1}{N_C}\,
 \frac{g_Z^6 ( g_{qL}^2 + g_{qR}^2 )}{4(\hat s- m_Z^2 )^2 } \, \left(\hat{s} 
  + \frac{4 p_1 \cdot q_1  \, p_2 \cdot q_1 }{m_Z^2} \right )
\end{equation}
where $N_C=3$ is the color factor of the quark, 
$g_{qL} = T_{3q} - Q_q \sin^2\theta_{\rm W}$,
      $g_{qR} =        - Q_q \sin^2\theta_{\rm W}$, and $\hat s = (p_1+p_2)^2$.
The differential cross section is
\begin{equation}
\label{eq:sigma}
d\sigma=  \frac{{\cal S}}{ (2\pi)^5 2 \hat s} 
\; \overline{\left| M \right |^2}
\; d PS_3 
\end{equation}
where $dPS_3$ is the 3-body phase space factor and the symmetric
factor ${\cal S} =1/2$ is due to a  pair of identical
Higgs bosons
in the final state.

We can consider both leptonic and hadronic decays of the $Z$ boson. The 
signature in the final state is a reconstructed $Z$ boson with a large
missing transverse energy. The irreducible SM background consists of
$ZZ$ production with one of the $Z$ bosons decaying into neutrinos. The 
other reducible backgrounds include $Z+$jets with the jets lost in the beam
pipe, $t\bar t$ production with the charged leptons or jets reconstructed 
at the $Z$ mass while all other particles are undetected. 
Since the irreducible background is already much larger than our signal, we only include it in our analysis.

The second subprocess that we consider at the LHC  is 
$q\bar{q} \to W HH$.
The spin- and color-averaged matrix element squared is given by
\begin{equation}
\overline{\left| M \right |^2 }
(q \bar q' \to WHH)
= \frac{1}{4}\,\frac{1}{N_C}\,
  \,
 \frac{ g^6}{8 (\hat s- m_W^2 )^2} \, \left( \hat s
  + \frac{4 p_1 \cdot q_1  \, p_2 \cdot q_1 }{m_W^2} \right ).
\end{equation}
The differential cross section is that in Eq.(\ref{eq:sigma}).
We consider both the leptonic and hadronic decays of the $W$ boson. The final
state consists of a reconstructed $W$ boson in the hadronic mode or an 
isolated charged lepton in the leptonic mode, 
plus large missing 
transverse energy. Irreducible background consists of $WZ$ production 
with the $Z$ boson decaying into neutrinos.

\begin{figure}[t!]
\centering
\includegraphics[width=3.2in]{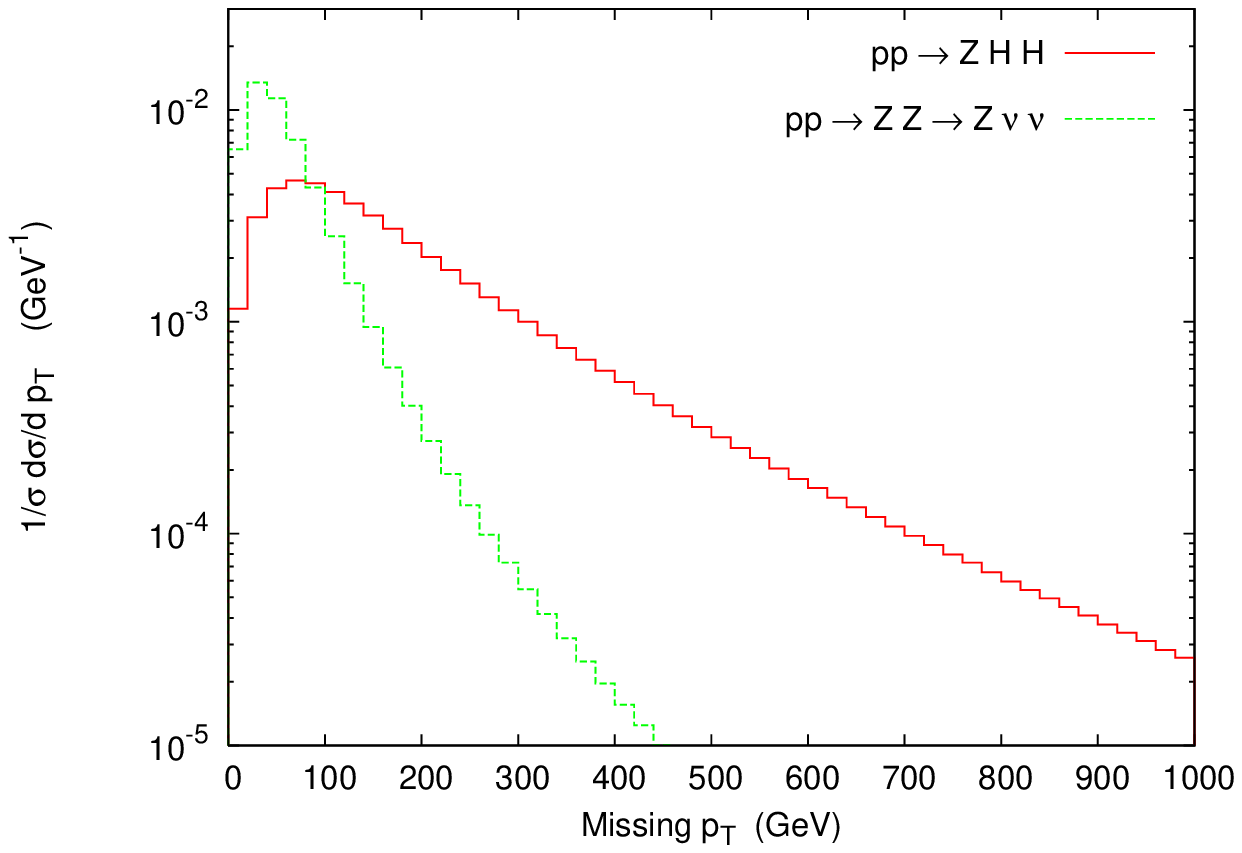}
\includegraphics[width=3.2in]{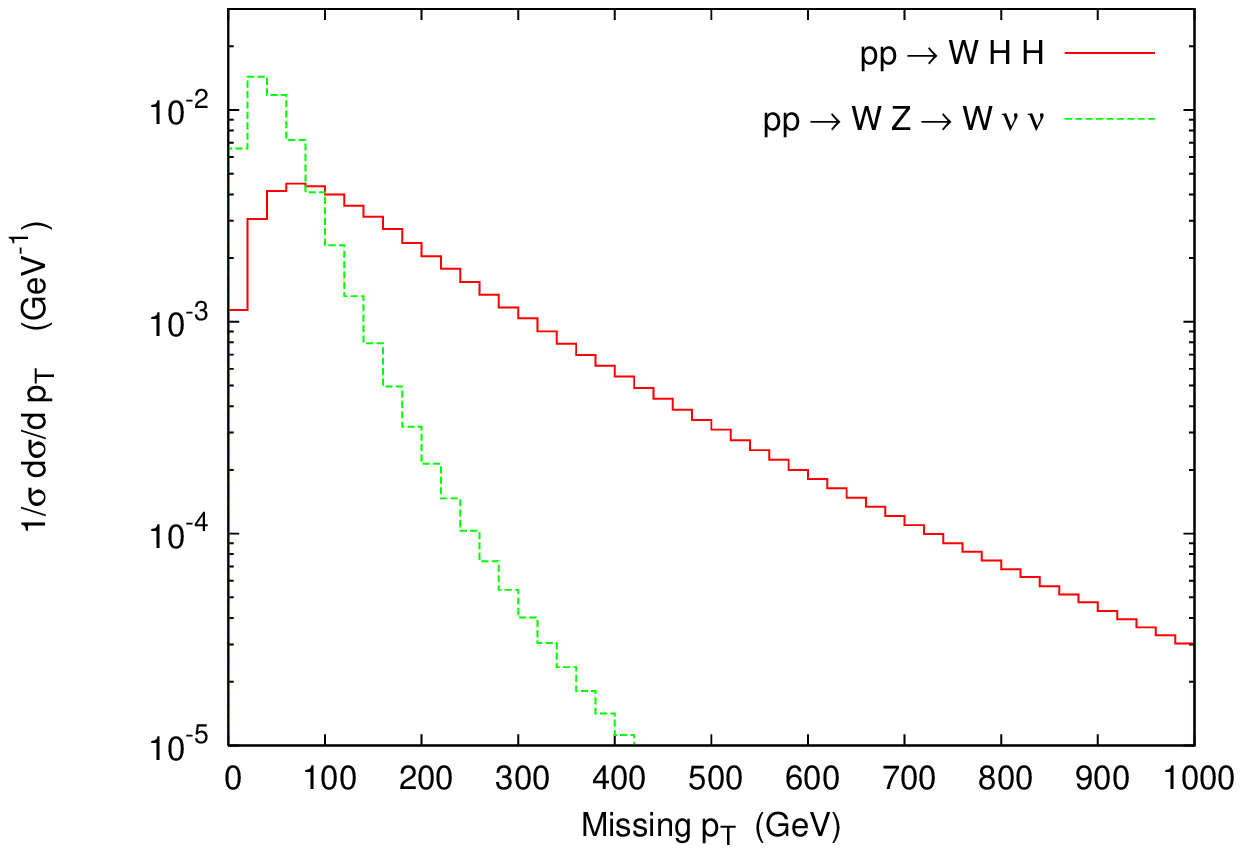}
\caption{\label{spectrum}\small
The {\it normalized} 
missing $p_T$ spectrum for (a) $ZHH$ and (b) $W^\pm HH$ production
at the LHC, with the corresponding background $ZZ \to Z \nu\bar \nu$ 
and $WZ \to W \nu \bar \nu$. The impose cut is $| y(W/Z) | < 2$.
}
\end{figure}

In Figs.~\ref{spectrum}
(a) and (b), we compared, with the corresponding backgrounds,
the spectrum  of the missing transverse 
momentum for the $ZHH$ and 
$WHH$ signals, respectively.
We have assumed that the $Z$ or the $W$ boson can be reconstructed.
The signal and the background have been
normalized, since the background is 3 orders of magnitude larger
than the signal. Applying a strong cut on the missing $p_T > 100\gev$, we arrive
at the signal-background ratio shown in Table \ref{table-lhc}. 
Larger $p_T$ cut would further hurt our signal.
Even with a possible luminosity of 100 fb$^{-1}$ the signal-background
significance is still very low to afford any positive identification
of the model.

\begin{table}[th!]
\caption{\label{table-lhc}
The cross section in fb for the signals $ZHH$ and $W^\pm HH$ and the
corresponding backgrounds $ZZ \to Z \nu\bar \nu$ and 
$WZ \to W \nu \bar \nu$ at the LHC.  The applied cuts include
$|y(Z/W) | < 2 $ and $\not\! {p}_T > 100$ GeV.
}
\centering
\begin{ruledtabular}
\begin{tabular}{ll|ll}
~~~~~  $ZHH$  &  $ZZ \to Z \nu\bar \nu$ ~~~~~  & ~~~~~   
$W^\pm HH$ &  $WZ \to W \nu \bar \nu$ ~~~~~ \\ \hline
~~~~~     0.2 fb & 370 fb~~~~~   & ~~~~~  0.4 fb & 390 fb 
 ~~~~~ \\
\end{tabular}
\end{ruledtabular}
\end{table}

\section{Signatures at the ILC}

\begin{figure}[t!]
\centering
\includegraphics[width=5in]{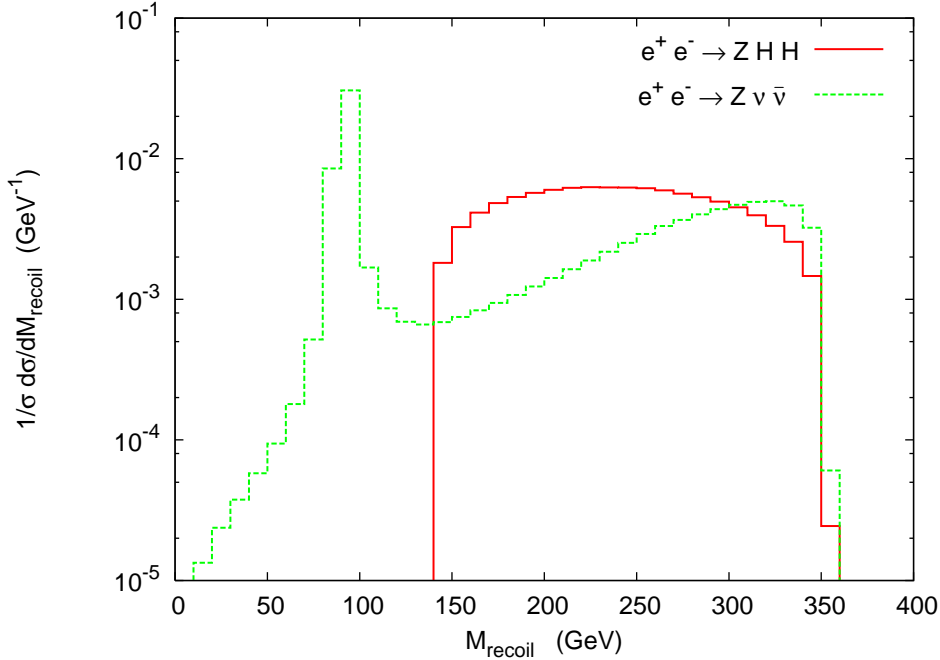}
\caption{\label{ee-fig} \small 
Normalized recoil mass spectrum for $e^- e^+ \to Z H H$ and 
$e^- e^+ \to Z \nu \bar \nu$, which includes three flavors of neutrinos,
at $\sqrt{s} = 0.5$ TeV.  Imposed cuts include
$|\cos \theta_Z | < 0.8$ and $p_{TZ} > 100$ GeV.
}
\end{figure}

The advantages of the ILC with electron
and positron beams include (i) known initial energy in the collision level,
and (ii) capability of polarization in the electron and positron beams.
With known initial energy we can calculate the mass of any particle(s) 
missing by measuring the energy of the visible 
particles---the method of recoil mass.  In the signal process
$e^- e^+ \to Z H H $, the $Z$ boson is measured 
by the momenta of the two charged leptons or the two jets 
while the two Higgs bosons are missing. We construct the
recoil mass by
\begin{equation}
m_{\rm recoil} = m_{HH} = \left[ s + m_Z^2 - 2 \sqrt{s} E_Z \right ]^{1/2} \;,
\end{equation}
where $m_{HH}$ is the invariant mass of the $HH$ system. 
Thus, the recoil mass spectrum will start at $2 m_H$ 
and be continuous with no peak
structure.  On the one hand, the recoil mass spectrum of 
the background process
$e^- e^+ \to \sum_{i=e,\mu,\tau}Z \bar\nu_i\nu_i$ is
very different.
For $Z\bar\nu_\mu\nu_\mu$ and $Z\bar\nu_\tau\nu_\tau$,
the dominant production mechanism is via
$e^- e^+ \to Z Z \to Z \bar\nu\nu$:
a sharp peak in the recoil mass distribution
emerges right at the mass $m_Z$.
Another mechanism is the $Z$ boson radiated off a leg
in $e^+ e^- \to \bar{\nu}\nu$, but it is very small.
However, $Z \bar{\nu}_e \nu_e$ production has additional
major production mechanism mediated by a $t$-channel $W$ boson.
This yields a continuous spectrum 
as well as a broad peak toward the high end
in the $m_{\rm recoil}$
distribution.
We show the normalized recoil mass spectrum in Fig. \ref{ee-fig}.

It is clear in Fig.\,\ref{ee-fig} that we
can use a recoil mass cut to remove the background due to the $Z$ peak.
However, the continuous SM background 
with the broad peak toward high end is 
still very large, thus reducing the sensitivity of our signal.
Since this continuous background mainly comes from 
the Feynman diagrams with a $W$ boson exchange, we
can use right-hand polarized electron beam and/or left-hand polarized
positron beam to largely reduce the background.  The remaining 
contribution comes from those with $Z$ boson exchange.
On the other hand, the signal receives 
compatible contributions from left-handed 
and right-handed electrons according to the size of 
$(g_{eL})^2\approx (-0.27)^2$ and $(g_{eR})^2\approx (0.23)^2$, respectively. 
In Table \ref{pol-x}, we show various cross sections
of the signal and the corresponding background, using cuts and 
realistic polarized beams.  Note that both signal and background
have zero cross sections when the electron and positron have
the same helicities, $(-,-)$ and $(+,+)$.

\begin{table}[th!]
\caption{\label{pol-x}
The cross section in fb for the signals 
$e^- e^+ \to Z^{\rm (vis)}HH$ and 
the corresponding backgrounds 
$e^- e^+ \to Z^{\rm (vis)} \nu\bar \nu$ 
at a 0.5 TeV ILC, using polarized beams.
Here $Z^{\rm (vis)}$ denotes the $Z$ boson 
considering only its visible decay 
(a branching ratio of 0.8 has been multiplied).
Imposed cuts include $| \cos \theta_Z | < 0.8$,
$p_{TZ} > 100$ GeV, and $m_{\rm recoil} > 140$ GeV.
The significance $S/\sqrt{B}$ is based on the ILC luminosity of
1000 fb$^{-1}$.
}
\centering
\begin{ruledtabular}
\begin{tabular}{|rr|rr|l|}
$P_{e^-}$  & $P_{e^+}$~~~~ & 
$\sigma(e^- e^+ \to Z^{\rm (vis)} \nu\bar \nu)$~~~~~ & 
$\sigma(e^- e^+ \to Z^{\rm (vis)}HH)$~~~~~ 
& $\dfrac{S}{\sqrt{B}}$ \\ 
\hline
$+1$ & $-1$~~~~~ & 3.8 fb ~~~~~ & 0.14 fb ~~~~~ & 2.2 ~~~~~ \\
$-1$ & $+1$~~~~~ & 200 fb ~~~~~ & 0.18 fb ~~~~~ & 0.4 ~~~~~ \\
0 & 0~~~~~ & 52 fb ~~~~~ & 0.08 fb ~~~~~ & 0.3 ~~~~~ \\
$0.8$ & $-0.6$~~~~~ & 6.8 fb ~~~~~ & 0.10 fb ~~~~~ & 1.2 ~~~~~ \\
\end{tabular}
\end{ruledtabular}
\end{table}

\section{Conclusions}

In the $SO(5) \times U(1)$ gauge-Higgs unification model
based on the Randall-Sundrum warped spacetime,
the Higgs boson is a part of the fifth component
of a higher-dimensional gauge field.
Because of the invariance of the 
effective Lagrangian under $H\to -H$,
this model accommodates the $H$ parity,
under which the Higgs field is odd while all the other SM 
particles are even.
The triple vertices of $HWW$, $HZZ$, and $H\bar{f}f$
vanish.
The very SM Higgs boson is the dark matter candidate.
At low energy, this model is highly constrained 
as it has only one free parameter,
the mass of the Higgs boson.  
The WMAP data on the relic density constrain $m_H\simeq 70\gev$.

Collider production of the Higgs boson must come in pairs. The most direct
way to detect is the associated production with a $W$ or $Z$ boson.
The collider signature is the $W/Z$ boson with a large $p_T$ missing.
We have shown that at the LHC the SM production of $WZ$ and $ZZ$ 
simply overwhelms
the signal even with quite strong kinematic cuts. 
Although the Higgs boson could also be pair produced 
in $WW$ fusion or $gg$ fusion, the detection of the merely
missing energy signal would be much more difficult.

On the other hand, we have shown that the whole situation substantially
improves at the ILC with polarized beams. 
The kinematic cut on the recoil mass can suppress the 
$e^- e^+ \to ZZ \to Z \nu \bar \nu$ background, and
a right-hand polarized electron
beam can dramatically reduce the $W$-mediated 
$e^- e^+ \to Z \bar \nu_e\nu_e$ background.  
We can therefore do an
event counting to measure the significance level of the signal.

\section*{Acknowledgments}
The work of K.C. was supported in part by the NSC of Taiwan and the WCU program through the KOSEF funded by the MEST
(R31-2008-000-10057-0).
The work of J.S. was supported by a Korean Research Foundation
grant (KFR-2009-013-C00014).

\end{document}